\documentclass{SciPost}

\hypersetup{
    colorlinks,
    linkcolor={red!50!black},
    citecolor={blue!50!black},
    urlcolor={blue!80!black}
}

\usepackage[bitstream-charter]{mathdesign}
\usepackage{subcaption}
\DeclareSymbolFont{usualmathcal}{OMS}{cmsy}{m}{n}
\DeclareSymbolFontAlphabet{\mathcal}{usualmathcal}

\fancypagestyle{SPstyle}{
\fancyhf{}
\lhead{\colorbox{scipostdeepblue}{\bf \color{white} ~SciPost Physics Proceedings }}
\rhead{{\bf \color{scipostdeepblue} ~Submission }}

\fancyfoot[C]{\textbf{\thepage}}
}

\begin{document}

\pagestyle{SPstyle}

\begin{center}{\Large \textbf{\color{scipostdeepblue}{
Automatizing the search for mass resonances using BumpNet\\
}}}\end{center}

\begin{center}


Jean-Fran\c{c}ois Arguin\textsuperscript{1},
Georges Azuelos\textsuperscript{1,2},
Émile Baril\textsuperscript{1},
Ilan Bessudo\textsuperscript{3},
Fannie Bilodeau\textsuperscript{1},
Maryna Borysova\textsuperscript{3},
Shikma Bressler\textsuperscript{3},
Samuel Calvet\textsuperscript{4},
Julien Donini\textsuperscript{4},
Etienne Dreyer\textsuperscript{3},
Michael Kwok Lam Chu\textsuperscript{3},
Eva Mayer\textsuperscript{4},
Ethan Meszaros\textsuperscript{1$\star$},
Nilotpal Kakati\textsuperscript{3},
Bruna Pascual Dias\textsuperscript{4},
Joséphine Potdevin\textsuperscript{1,5},
Amit Shkuri\textsuperscript{3} and
Muhammad Usman\textsuperscript{1}

\end{center}

\begin{center}
\small
{\bf 1} Group of Particle Physics, Universit\'e de Montr\'eal, Montr\'eal QC; Canada \\
{\bf 2} TRIUMF, Vancouver BC; Canada \\
{\bf 3} Department of Particle Physics and Astrophysics, Weizmann Institute of Science, Rehovot; Israel \\
{\bf 4} LPCA, Universit\'e Clermont Auvergne, CNRS/IN2P3, Clermont-Ferrand; France \\
{\bf 5} Institute of Physics, Ecole Polytechnique F\'ed\'erale de Lausanne (EPFL), Lausanne; Switzerland

\vspace{0.5em}
%
$\star\:$\href{mailto:ethan.james.meszaros@cern.ch}{\small ethan.james.meszaros@cern.ch}

\end{center}

\definecolor{palegray}{gray}{0.95}
\begin{center}
\colorbox{palegray}{
  \begin{tabular}{rr}
  \begin{minipage}{0.37\textwidth}
    \includegraphics[width=60mm]{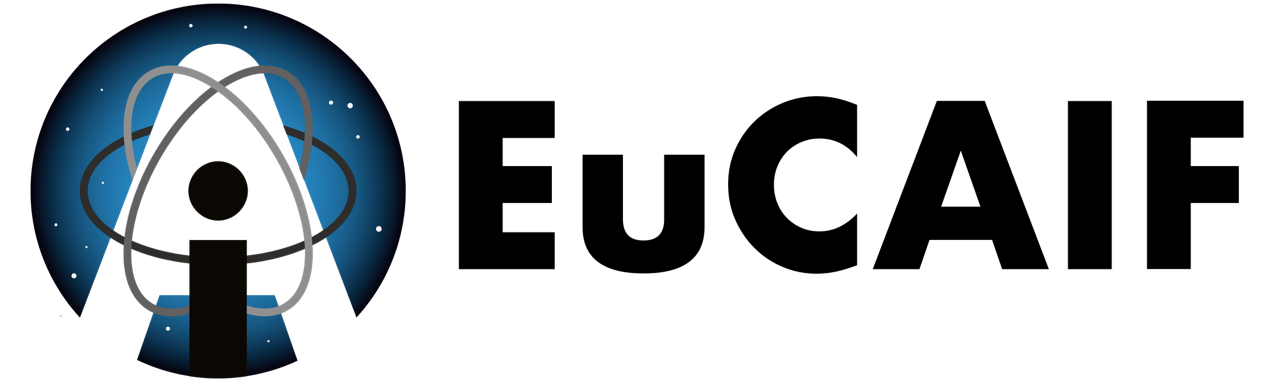}
  \end{minipage}
  &
  \begin{minipage}{0.5\textwidth}
    \vspace{5pt}
    \vspace{0.5\baselineskip} 
    \begin{center} \hspace{5pt}
    {\it The 2nd European AI for Fundamental \\Physics Conference (EuCAIFCon2025)} \\
    {\it Cagliari, Sardinia, 16-20 June 2025
    }
    \vspace{0.5\baselineskip} 
    \vspace{5pt}
    \end{center}
    
  \end{minipage}
\end{tabular}
}
\end{center}

\section*{\color{scipostdeepblue}{Abstract}}
\textbf{\boldmath{%
Physics Beyond the Standard Model (BSM) has yet to be observed at the Large Hadron Collider (LHC), motivating the development of model-agnostic, machine learning-based strategies to probe more regions of the phase space. As many final states have not yet been examined for mass resonances, an accelerated approach to bump-hunting is desirable. BumpNet is a neural network trained to map smoothly falling invariant-mass histogram data to statistical significance values. It provides a unique, automatized approach to mass resonance searches with the capacity to scan hundreds of final states reliably and efficiently. 
}}

\vspace{\baselineskip}

\noindent\textcolor{white!90!black}{%
\fbox{\parbox{0.975\linewidth}{%
\textcolor{white!40!black}{\begin{tabular}{lr}%
  \begin{minipage}{0.6\textwidth}%
    {\small Copyright attribution to authors. \newline
    This work is a submission to SciPost Phys. Proc. \newline
    License information to appear upon publication. \newline
    Publication information to appear upon publication.}
  \end{minipage} & \begin{minipage}{0.4\textwidth}
    {\small Received Date \newline Accepted Date \newline Published Date}%
  \end{minipage}
\end{tabular}}
}}
}




\section{Introduction}
\label{sec:intro}
Though the Standard Model (SM) is known to be incomplete \cite{weinberg_essay_2018}, signatures predicted by Beyond the Standard Model (BSM) extensions have yet to be identified. Over the last decade, many model-agnostic, machine learning-fueled techniques have been developed for new physics discovery with the intent of broadening signal sensitivity and accelerating analyses. However, many of these methods require some form of traditional background estimation and are limited by training sample sizes, low signal fractions in data, and validation issues \cite{belis_machine_2024}. 

Historically, ``bump hunting'' has been effective at finding new particles (see Refs.~\cite{herb_observation_1977, aubert_experimental_1974, aad_observation_2012}) due to the model-agnostic nature of mass resonances. Given that many observable final states at the LHC have yet to be examined \cite{kim_motivation_2020}, a broad search for resonances is well motivated. 

\textit{BumpNet} \cite{arguin_automatizing_2025} is a fully supervised technique for conducting resonance searches efficiently across hundreds of final states. It utilizes a well-established property of the SM, that of smoothly falling backgrounds, to map invariant-mass distributions to local statistical significance values. This frees BumpNet from the need for prior background estimation and signal assumptions. The following sections detail its architecture, training procedure, and performance in a proof-of-concept study.

\begin{figure}
    \centering
    \includegraphics[width=0.65\linewidth]{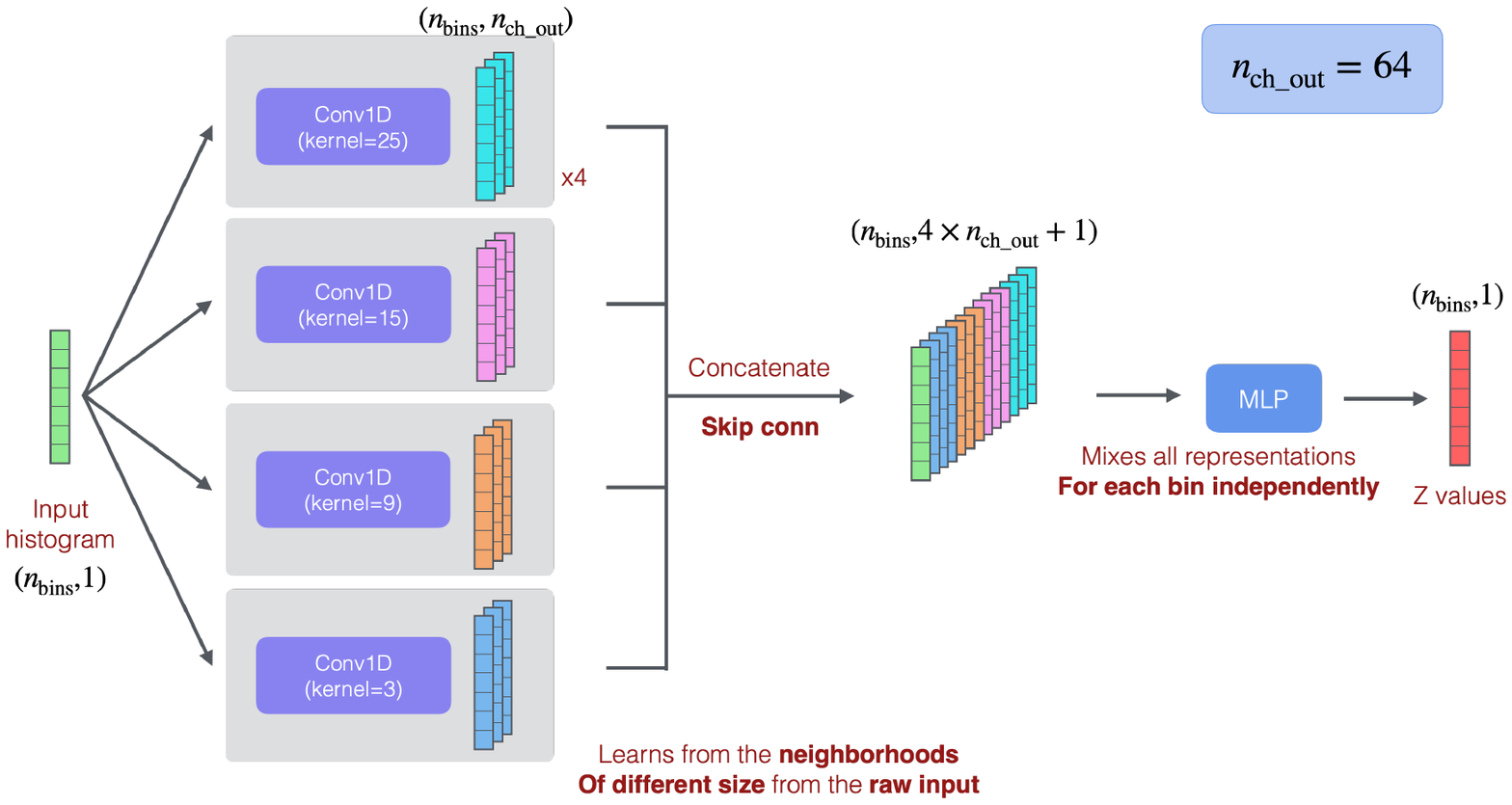}
    \vspace{-2em}
    \caption{BumpNet's architecture, receiving histogram entries as inputs and outputting the predicted local significance in each bin.}
    \label{fig:architecture}
\end{figure}

\section{Methodology}
\label{sec:methodology}

\subsection{Architecture}
BumpNet's architecture is convolution-based, as seen in Figure \ref{fig:architecture}. It is designed to receive as input the number of events in each bin of smoothly falling invariant-mass histograms. It receives no information about the invariant-mass itself, making BumpNet resistant to potential invariant-mass correlations. The input is fed into four convolutional stacks, each with a unique kernel size to capture different sized patterns in the input. This convolved representation is fed, bin-by-bin, into a multilayer perceptron (MLP) which outputs a single value: the prediction of the local significance in a given bin. This is repeated to obtain the significance prediction across the entire histogram range. 

\subsection{Training}
The training dataset is created using an assortment of smoothly falling backgrounds modeled by (1) a group of eleven analytical functions, with parameters constrained to satisfy a certain dynamic range, and (2) the "smoothing" of monte carlo (MC) distributions. The latter method is introduced to account for the fact that analytical functions do not necessarily match the shapes seen in real data. These MC distributions used for smoothing are produced through a histogram production framework, described in Section \ref{sec:HistogramProduction}, then subsequently smoothed by parametric and non-parametric fitting methods to extract shapes that more closely resemble those expected to be seen in real data. 

Once a sample of smoothly falling curves has been created, a 1 bin-width wide gaussian signal is injected at random positions and with random signal strengths. These final curves are Poisson-fluctuated to resemble realistic distributions, and the local significance of deviations from the known background (calculated using formulae from Ref. \cite{cowan_asymptotic_2011}) serve as the network's labels.


\subsubsection{Histogram Production}\label{sec:HistogramProduction}
To demonstrate BumpNet's methodology, distributions for smoothing are generated using the Dark Machines (DM) MC dataset \cite{aarrestad_dark_2022}. This dataset emulates $10 \: \text{fb}^{-1}$ of SM $pp$ collisions at $\sqrt{s}=13$ TeV, including the 26 highest cross-section processes expected at the LHC. Exclusive final states are constructed from all combinations of the available objects, including electrons, muons, photons, jets, $E_T^{miss}$ and some specially defined objects, such as a boosted hadronic $W/Z$. One invariant-mass histogram is then created for every combination of at least two objects within a given final state, yielding 39,768 total invariant-mass histograms.

Bin sizes vary according to an approximate mass resolution in order to make resonances appear similar to BumpNet in units of bins. A similar histogram production framework will be used to generate histograms from real data in future analyses.

\section{Performance}
\label{sec:performance}
For the DM proof-of-concept \cite{arguin_automatizing_2025}, BumpNet is trained on approximately 3 million samples, one third of which are background shapes obtained from analytical functions and the remaining from smoothed MC histograms. To examine performance, BumpNet is tested on an application set generated in the same manner as its training dataset and on various BSM scenarios. 

\begin{figure}[ht]
    \centering
    \begin{minipage}{\textwidth}
        \centering
        \begin{subfigure}[t]{0.32\linewidth}
            \centering
            \includegraphics[width=\linewidth]{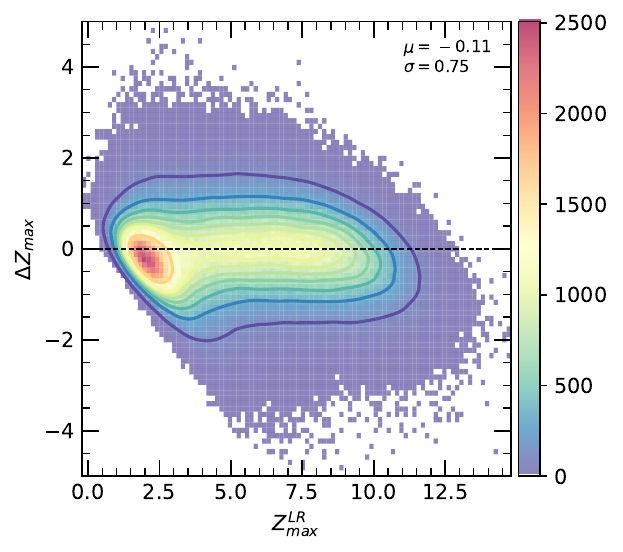}
            \caption{$\Delta Z_{max}$ v. $Z_{max}^{LR}$}
            \label{fig:deltazmax_v_zlrmax}
        \end{subfigure}
        \hfill 
        \begin{subfigure}[t]{0.32\linewidth}
            \centering
            \includegraphics[width=\linewidth]{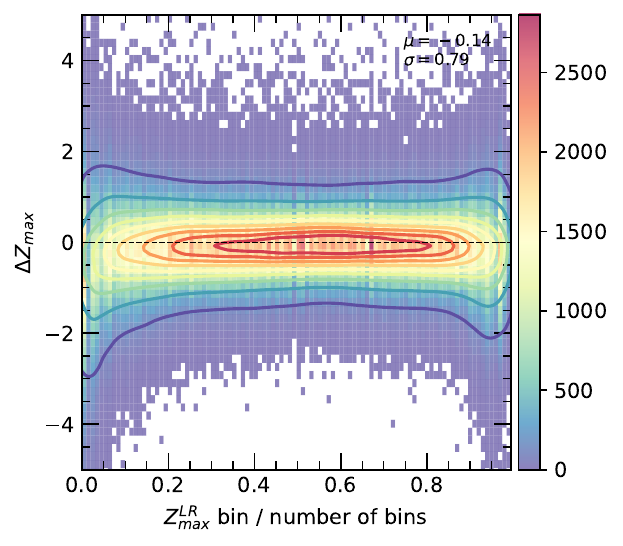}
            \caption{$\Delta Z_{max}$ v. relative position of $Z_{max}^{LR}$ (includes all bins)}
            \label{fig:deltazmax_v_ratio}
        \end{subfigure}
        \hfill 
        \begin{subfigure}[t]{0.32\linewidth}
            \centering
            \includegraphics[width=\linewidth]{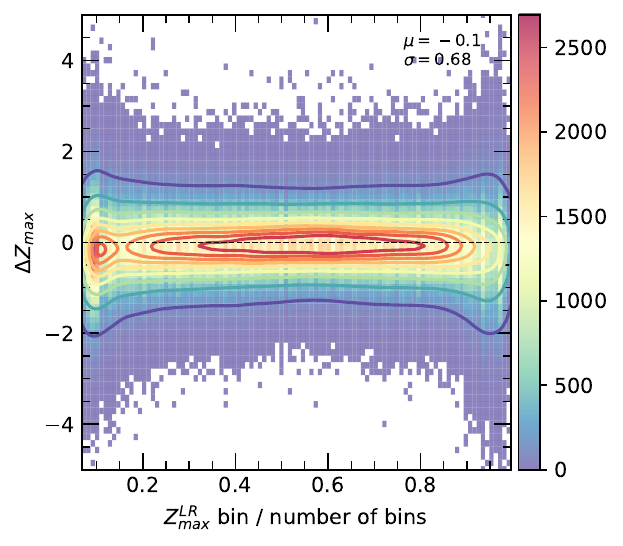}
            \caption{$\Delta Z_{max}$ v. relative position of $Z_{max}^{LR}$ (first $10\%$ of bins excluded)}
            \label{fig:deltazmax_v_ratio_10per}
        \end{subfigure}
        \caption{BumpNet's agreement with $Z_{max}^{LR}$ measured as a function of signal strength (\ref{fig:deltazmax_v_zlrmax}) and relative position (\ref{fig:deltazmax_v_ratio}, \ref{fig:deltazmax_v_ratio_10per}).}
        \label{fig:deltazmax}
    \end{minipage}
\end{figure}

\subsection{Injected Gaussian Signals}
A total of 500,000 function- and 1,000,000 DM-based histograms are used for the application set. Performance is assessed by calculating $\Delta Z_{max} = Z_{max}^{BumpNet} - Z_{max}^{LR}$, the difference between BumpNet's prediction and the true significance, respectively. Figure \ref{fig:deltazmax} shows that, as a function of various signal strengths and positions, $\Delta Z_{max}$ is unbiased with relatively small spread. Excluding the predictions in the first $10\%$ of bins reduces the spread further (Figure \ref{fig:deltazmax_v_ratio_10per}), emphasizing the ambiguity in the inferred background near the beginning of the histograms. 


\subsection{Realistic Signals and the Look Elsewhere Effect}
\begin{figure}[ht]
    \centering
    \begin{minipage}{\textwidth}
        \centering
        \begin{subfigure}{0.45\linewidth}
            \centering
            \includegraphics[width=0.8\linewidth]{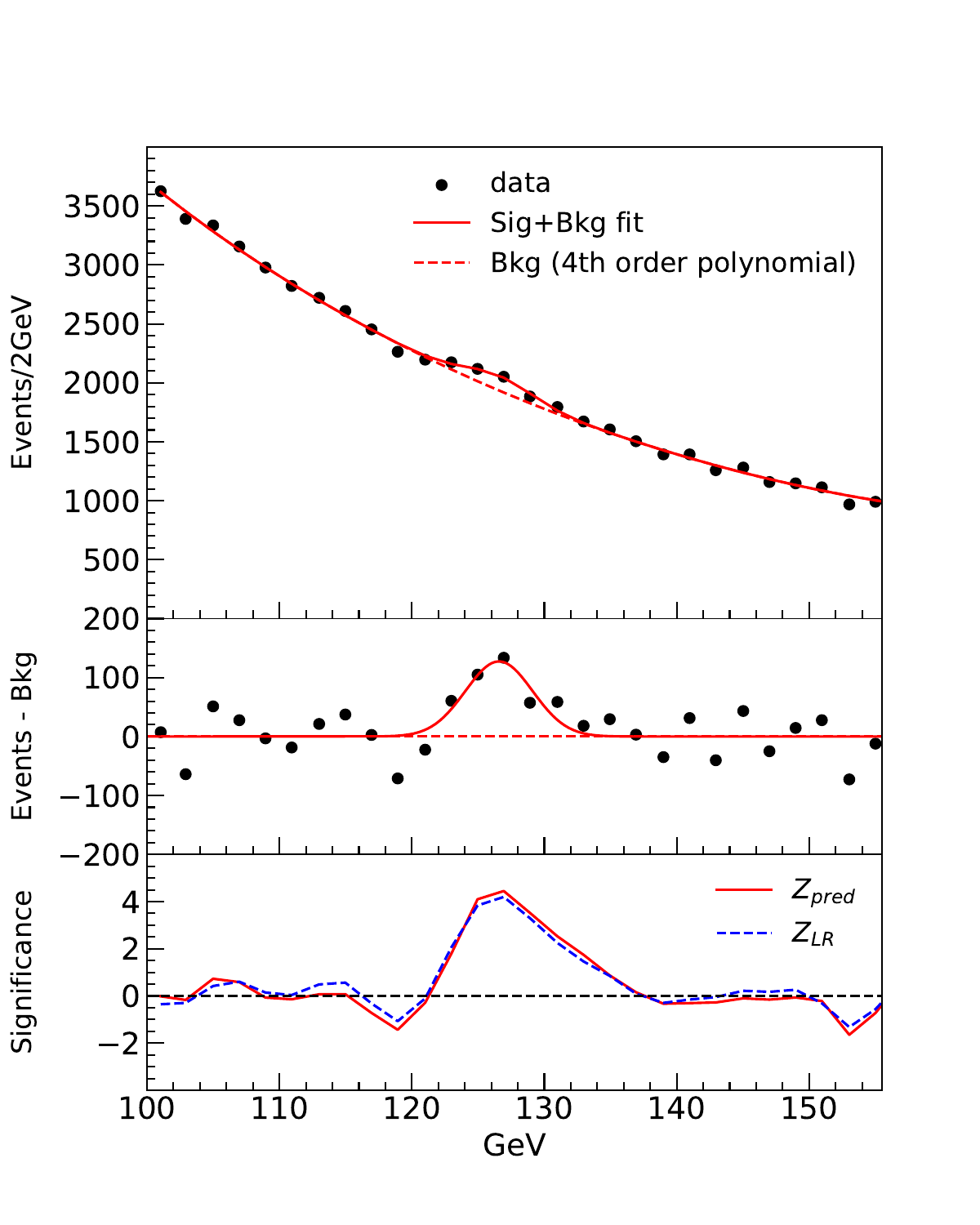}
            \caption{Higgs discovery data}
            \label{fig:higgs}
        \end{subfigure}
        \begin{subfigure}{0.5\linewidth}
            \centering
            \includegraphics[width=0.8\linewidth]{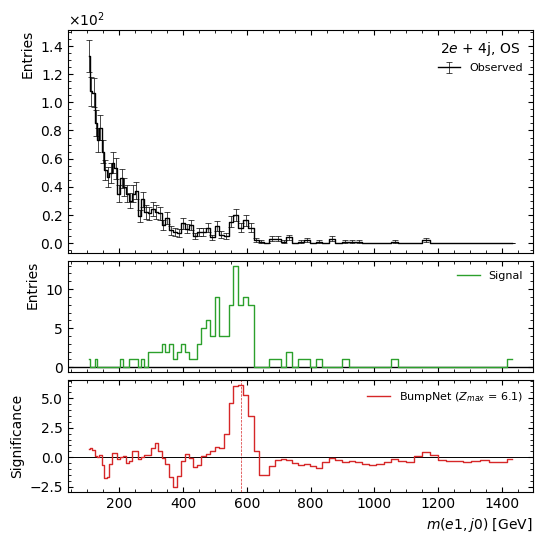}
            \caption{Two $LQ \rightarrow be$}
            \label{fig:LQ}
        \end{subfigure}
        \caption{Examples of BumpNet's performance over data-like signals. In \ref{fig:higgs}, the maximum prediction ($Z_{max}^{pred}$) of $4.5\sigma$ agrees with the likelihood-ratio calculation ($Z_{max}^{LR}$) of $4.2\sigma$ within BumpNet's uncertainties. In \ref{fig:LQ}, the prediction peaks where the most BSM events are located, demonstrating BumpNet's sensitivity to general signatures. }
        \label{fig:data-like-signals}
    \end{minipage}
\end{figure}

To test sensitivity to realistic physics scenarios, BumpNet is applied on previously analyzed ATLAS data (see Figure \ref{fig:higgs} and Section 4.1 of Ref. \cite{arguin_automatizing_2025}) and BSM samples injected on top of SM background. These BSM samples include both locally generated events and those provided in the DM dataset. The network performs well across multiple scenarios, including the pair production of scalar leptoquarks with a mass of 600 GeV depicted in Figure \ref{fig:LQ}. 

Despite BumpNet's stellar performance, the natural problem of false positive signals and the look elsewhere effect (LEE) arises swiftly when scanning thousands of histograms for bumps. To mitigate such false positives, an algorithm exploiting physical correlations between object combinations (described in Section 4.2 of Ref. \cite{arguin_automatizing_2025}) has been developed to much success. A framework has also been deployed for calculating the global significance of BumpNet's predictions, which will strengthen the results of future analyses by quantifying the extent of the LEE. 

\vspace{-2em}

\section{Conclusion}
The BumpNet methodology presents a promising approach to automatizing the mass resonance search, enabling an efficient scan of hundreds of final states to indicate areas of interest. The performance of our proof-of-concept model is unbiased with low variance and demonstrates sensitivity to a wide variety of possible physics scenarios. Multiple approaches have been developed to mitigate the inevitable look elsewhere effect, highlighting BumpNet's potential to reliably scan more regions of the phase space than any prior analysis. 

\section*{Acknowledgements}


\paragraph{Funding information}

We would like to acknowledge the support provided by the Natural Sciences and Engineering Research Council of Canada (NSERC), the Institut de valorisation des données (IVADO), the Canada First Research Excellence Fund, and the Israeli Science Foundation (ISF, Grant No. 2382/24). We are also grateful to the Krenter-Perinot Center for High-Energy Particle Physics, the Shimon and Golde Picker-Weizmann Annual Grant, and the Sir Charles Clore Prize for their generous support. We further wish to thank Martin Kushner Schnur for his significant and invaluable contribution to this work.

\bibliography{bibliography.bib}


\end{document}